\documentstyle[11pt,newpasp,twoside,epsf]{article}
\def\edcomment#1{\iffalse\marginpar{\raggedright\sl#1\/}\else\relax\fi}
\reversemarginpar

\begin{document}

\newcommand{\kms}{\,km\,s$^{-1}$}     
\def\osix{O\,{\sc vi}}
\def\cfour{C\,{\sc iv}}
\def\hone{H\,{\sc i}}
\def\ctwo{C\,{\sc ii}}

\title{O\,{\sc vi} High Velocity Clouds: Where and What are They?}
\author{Kenneth R. Sembach}
\affil{Department of Physics \& Astronomy, The Johns Hopkins University,
3400 N. Charles Street, Baltimore, MD  21218}

% Add as many "\author" and "\affil" fields as necessary.

\begin{abstract}
Our view of the high velocity cloud (HVC) system is changing 
dramatically with new observations from the {\it Far Ultraviolet Spectroscopic 
Explorer} mission and the Space Telescope Imaging Spectrograph on-board
the {\it Hubble Space Telescope}. In particular, 
the detection of \osix\ absorption indicates that many HVCs have a hot, 
collisionally-ionized component in addition to the neutral gas detected 
in \hone\ 21\,cm emission.  A better understanding of the ionization of this 
gas will help to determine the distances of some HVCs, as well as constrain 
the properties of a hot Galactic corona or Local Group medium. 
In this article,  I briefly summarize some recent observations of \osix\ HVCs. 
\end{abstract}

\section{Introduction}
It is my sincere pleasure to be able to present a talk at this conference in 
honor of Dr. Ray Weymann and his many contributions to astronomy.  Rather 
than delve
into a lengthy description of the known properties of high velocity clouds 
(HVCs), I will instead concentrate on some recent observational results that 
provide new information about HVCs that you may not have heard about before.
I hope that this talk stimulates as much discussion about what we do not
yet know about HVCs as it does about what we do know. 

First, a very brief introduction to HVCs is in order. Despite their simple 
name, high velocity clouds have a long history of being contentious 
subjects of debate.  Most HVCs have been identified through their
\hone\ 21\,cm emission, which occurs at velocities well outside the range 
expected for typical interstellar clouds (typically, $|v_{HVC}| \sim 100-350$
\kms).  Within the last decade it has also been possible to study HVCs in 
absorption using ultraviolet spectrographs in space.  Some of these clouds 
have been confirmed to be Galactic in nature since they are observed in the 
absorption-line spectra of stars at known distances.  Others,
such as the Magellanic Stream, are clearly located outside the normally
recognized confines of the Galaxy.  The case for an extragalactic origin of 
some \hone\ HVCs has been presented recently by several authors (e.g.,
Blitz et al. 1999; Braun \& Burton 1999; see also Gibson et al. 
- this volume - for additional references and a concise discussion of the 
pros and cons of an extragalactic location for HVCs).  There are several 
comprehensive reviews of the sky distribution of HVCs and the known 
properties and locations of some of the better-studied clouds 
(Wakker \& van~Woerden 1997; Wakker 2001).

\section{Ionized Gas: A New Perspective}
Much of what is known about the kinematics, morphologies, and angular 
projections of HVCs onto the sky comes
from \hone\ 21\,cm emission studies.  However, many HVCs are now known
to contain significant quantities of ionized gas.  The ionized gas has been 
traced through both its H$\alpha$ emission (e.g., Bland-Hawthorn et al.
1998; Tufte, Reynolds, \& Haffner 1998) and its absorption-line signatures in
ionized species such as C\,{\sc iii-iv}, Si\,{\sc iii-iv}, and \osix\
(e.g., Sembach et al. 1999; Sembach et al. 2000).  

The existence of an ionized component to the high velocity cloud system has 
several obvious ramifications.  First, it indicates that there is material 
in a form that heretofore has not been readily observed.  The extent of this 
ionized gas is presently unknown, although new H$\alpha$ imaging techniques
hold great promise for revealing very low levels of emission regardless
of the distances of the clouds (e.g., Glazebrook \& Bland-Hawthorn 2001).
Second, the ionization of this gas must be maintained by an energy source
either internal or external to the HVCs.  Since there are as yet no known
stars that emit significant numbers of ionizing photons directly associated 
with HVCs, it seems likely that most of the energy must come either from 
photons emanating from somewhere outside the clouds or from collisional
processes occurring between the clouds and other interstellar or intergalactic
material.  Third,
the detection of high velocity \osix\ along many sight lines through the 
Milky Way halo (Sembach et al. 2000, 2001c; Savage et al. 2001) 
suggests that there may be dynamical interactions between some HVCs and 
a hot Galactic corona or Local Group medium.  It is to this third point that 
the rest of this article is devoted.

\section{\osix\ High Velocity Clouds}

The study of highly ionized high velocity gas traced through \osix\ 
absorption became possible with the commissioning of the {\it 
Far Ultraviolet Spectroscopic Explorer} ($FUSE$) in late 1999 (Moos et al. 
2000)\footnotemark.  Conversion of 
O\,{\sc v} into \osix\ requires $\sim114$~eV, an energy well above the 
He\,{\sc ii} absorption edge at 54.4~eV. \osix\ is an excellent tracer
of collisional processes at temperatures $T \sim (1-5)\times10^5$\,K
(Sutherland \& Dopita 1993).
\footnotetext{Additional information
about the general usefulness of $FUSE$ data for HVC studies can be found 
in Sembach (1999).}
\osix\ has an observable resonance doublet at 1031.926, 1037.617\,\AA.  Of 
the two lines, the stronger  (1031.926\,\AA) line is more frequently observed 
since the weaker line is often blended with other atomic
and molecular lines (e.g., \ctwo\ $\lambda1036.337$, 
\ctwo$^{*} \lambda1037.018$, H$_2$ (5--0) R(1) $\lambda1037.149$ and 
H$_2$ (5--0) P(1)
$\lambda1038.157$).  High velocity absorption in the \osix\ 1031.926\,\AA\ 
line can be blended with nearby H$_2$ 
(6--0) P(3) absorption at 1031.191\,\AA\ (--214 \kms) and (6--0) R(4)
absorption at 1032.349\,\AA\ (+122 \kms).  When a comparison between both 
members of the \osix\ doublet has been possible in the Milky Way halo and HVC
studies to date,  the results have indicated that there is no unresolved 
saturated structure in the lines at $FUSE$ resolution 
(FWHM~$\approx20-25$ \kms). This result is consistent with production of 
the \osix\ in collisionally-ionized gas at temperatures $T \ge 10^5$\,K. 

\begin{figure}
\plotfiddle{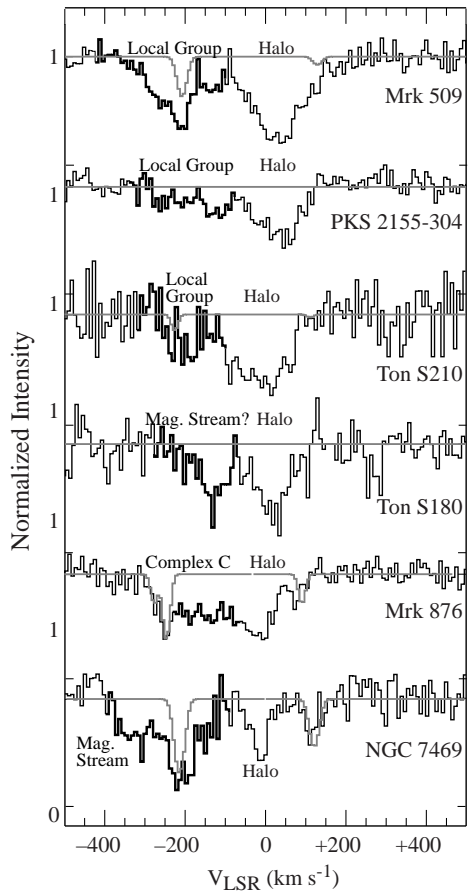}{4.5in}{0}{70}{70}{-200}{-99}
%\special{psfile=figure1.ps hoffset=-90 voffset=-600 vscale=80 hscale=80}
\caption{\scriptsize \osix\ HVCs observed by $FUSE$ along 6 sight lines
(from Sembach et al. 2000).  The \osix\ $\lambda1031.926$ data are shown,
together with a model of the Milky Way H$_2$ absorption expected in the 
(6--0) P(3) and R(4) lines (solid grey curves).  Milky Way disk and halo
absorption occurs along each sight line at velocities of about --100 to +100 
\kms.  The data have a velocity 
resolution (FWHM) of $20-25$ \kms.  A general location is indicated for each 
HVC delineated by the thick black lines.  All of these HVCs are located 
either in the very distant Galactic halo or outside the Galaxy.}
\end{figure}

\subsection{Examples}

Examples of the \osix\ $\lambda1031.926$ absorption in several 
high velocity clouds observed by $FUSE$ are 
shown in Figure~1.  These include the Magellanic Stream, Complex C, and HVCs
 believed to be located outside the Galaxy in the Local Group.  
Absorptions from both the Milky Way halo and the high velocity gas are 
evident.  The faint solid grey line overplotted on each spectrum is a model
of the H$_2$ absorption features expected in the velocity range shown (see 
Sembach et al. 2000).  The \osix\ HVCs in Figure~1 are relatively 
well-separated from the lower velocity absorption.  They occur at velocities 
similar to those of lower
ionization species (such as \cfour\ or \hone) and have strengths ranging 
from $\approx0.3$ (Ton\,S210) to $\approx1.7$ (NGC\,7469) times the strength 
of the Milky Way halo absorption.

A second type of high velocity gas detected in \osix\ absorption occurs
along some sight lines through the Milky Way halo.  Figure~2 shows an example 
of a high velocity \osix\ absorption wing in the direction of 3C\,273
(Sembach et al. 2001c).  This feature is shallow and very broad, roughly 
150 \kms\ in width.  The absorption is present in data obtained in multiple 
$FUSE$ 
channels and has no counterpart in lower ionization species.  The column of
\osix\ contained in this wing is approximately 10\% of the total \osix\
column along the sight line.  These high velocity wings are seen along several 
other sight lines currently under study and will be described in a
forthcoming paper.  The favored interpretation for the high velocity wing 
toward 3C\,273 is that the absorption traces the expulsion of hot gas out of 
the Galactic disk into the halo in the Loop~I / Loop~IV region of the sky.
The high ion ratios observed are consistent with the expectations for 
radiatively cooling gas in a fountain flow or supernova remnant
(see Sembach et al. 2001c).

\begin{figure}
\plotfiddle{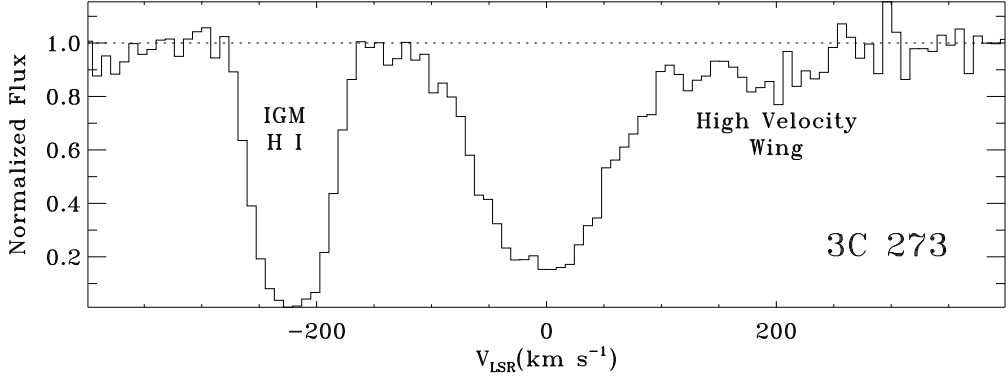}{2.2in}{0}{80}{80}{-230}{-240}
\caption{\footnotesize Galactic \osix\ $\lambda1031.926$ absorption toward 
3C\,273 (Sembach et al. 2001c).  Note the high velocity wing at $v_{LSR} > 
100$ \kms.  This absorption wing is observed only in \osix.  The high 
velocity \osix\ traces hot material flowing out of the Galactic disk into the 
halo. }
\end{figure}

\subsection{Relationship to \hone\ and Lower Ionization Species}

Most of the \osix\ HVCs studied so far have readily identifiable counterparts 
in \hone.  In the case of at least one large HVC, that of Complex C, 
the \osix\ absorption along several sight lines through the HVC 
traces the \hone\ velocity structure quite well.  Figure~3 shows the \osix\
$\lambda1031.926$ absorption toward four Complex~C sight lines together
 with the \hone\ 21\,cm emission profiles\footnotemark\ observed with the 
NRAO 140-foot telescope and the Fe\,{\sc ii} $\lambda1144.938$ profiles
observed with $FUSE$.  \footnotetext{These \hone\ 21\,cm data are from an 
unpublished survey by Murphy, Sembach, and Lockman.}  The \hone\ emission 
velocities and strengths change substantially across Complex~C.  For the 
four sight lines shown, the Complex~C \hone\ velocities range from 
$\approx-174$ \kms\ toward Mrk\,876 to $\approx-105$ \kms\ toward Mrk\,817.  
(This sight line 
pair has the smallest angular separation of the set, $\Delta \approx 
5.3\deg$.)  The \osix\ $\lambda1031.926$ 
absorption profiles clearly reveal gas at the Complex~C velocities
traced by \hone\ and Fe\,{\sc ii}.  The \osix\ profiles are smoother than 
the \hone\ emission or low ionization absorption, but track the neutral gas 
well; there appears to be a strong correlation between the \osix\ 
velocity structure and 
the velocity structure of lower ionization stages.  The \osix\ HVC line 
strength is roughly comparable along the Mrk\,817, Mrk\,279, and PG\,1259+593
sight lines, while the strength toward Mrk\,876 is about a factor of 
two higher.  The opposite is true for the HVC \hone\ emission; the strength 
of the emission toward Mrk\,876 is less than that along the other three sight 
lines.  

\begin{figure}
\plotfiddle{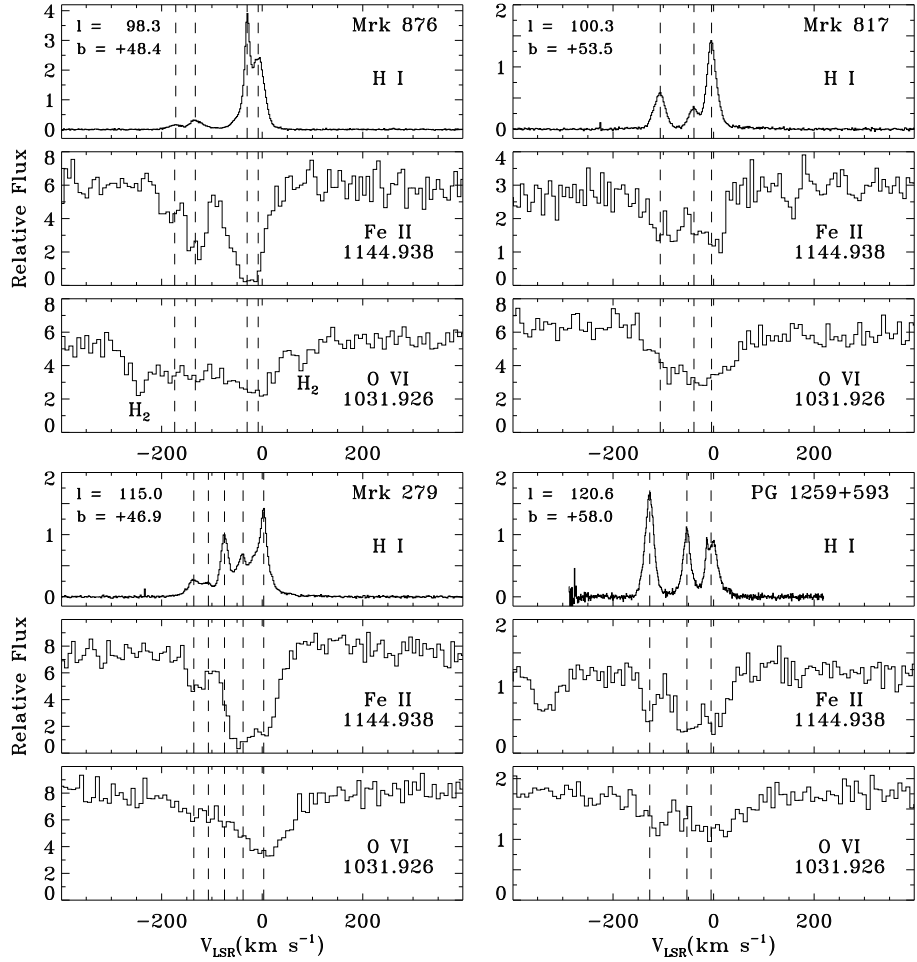}{4.8in}{0}{70}{70}{-220}{-150}
\caption{\footnotesize \osix\ $\lambda1031.926$ and Fe\,{\sc ii} 
$\lambda1144.938$ absorption 
toward four sight lines piercing through HVC Complex C.  Also shown are the 
H\,{\sc i} 21\,cm emission profiles observed along each sight line with
the NRAO 140-foot telescope.  
High and low ion Complex~C absorption features in the $FUSE$ spectra
at $v_{LSR} \le -100$ \kms\ are indicated by the left-most vertical lines.
Milky Way ISM absorption and emission are present near $v_{LSR} = 0$ \kms.
Differing amounts of intermediate velocity gas are
also present along the four sight lines.  Note that the bulk absorption
velocities of the neutral and \osix-bearing gases are similar, suggesting
that the highly ionized and neutral gases are in close proximity to each 
other.}
\end{figure}

We show a second comparison of the \osix\ and low ionization lines in another 
HVC in Figure~4.  This HVC, observed along the NGC\,1705 sight line 
($l~=~261.1\deg, b = -38.7\deg$), traces material associated with the 
Magellanic Stream near the LMC (see Sembach et al. 2001b).  This HVC has 
a multi-component structure as revealed by the Space Telescope Imaging 
Spectrograph data shown in Figure~4.  Multiple ionized gas components traced 
by Si\,{\sc iii} $\lambda1206.500$ are spread over a velocity range of 
$\approx$140 \kms.  The \osix\ absorption is centered near the velocity of the 
strongest low ionization component traced by the  O\,{\sc i} $\lambda1302.168$ 
line.  Furthermore, the \osix\ tracks closely the high velocity 
($v_{helio} > 320$ \kms) absorption wing seen Si\,{\sc iii}, suggesting either
that there is a hot component of the cloud centered on the strongest low
ionization component, or that there is a highly ionized leading edge to the 
cloud.  The width of the \osix\ absorption (FWHM $\approx 83$ \kms) is 
much broader than expected for a single-component
gas at $T \sim 3\times10^5$\,K, suggesting
that the \osix\  may trace multiple hot gas components or that the 
the absorption is broadened by non-thermal motions having velocities on the 
order of 50 \kms.

\begin{figure}
\plotfiddle{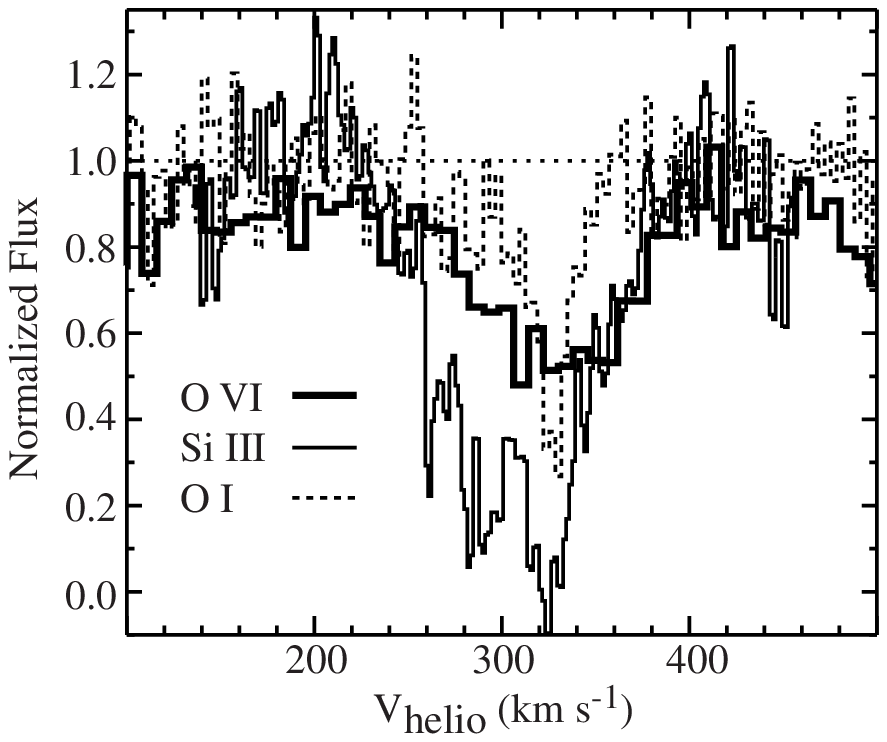}{2.2in}{0}{80}{80}{-230}{-240}
\caption{\footnotesize High velocity \osix\ $\lambda1031.926$,  
Si\,{\sc iii} $\lambda1206.500$, O\,{\sc i} $\lambda1302.168$ absorption 
in the HVCs toward NGC\,1705.  Note the multi-component structure of the 
ionized gas traced by Si\,{\sc iii}. The \osix\ absorption is roughly 
centered on the dominant neutral component traced by O\,{\sc i}.}
\end{figure}

\section{Where are the \osix\ HVCs?}

The answer to this question is still being sought.  However, from the 
information that is now available, it seems quite likely that {\it some} 
of the \osix\ HVCs are tracing material in the vicinity of the Milky Way 
rather than material within the Milky Way.  Several lines of evidence support 
this conclusion:  
\smallskip
\\
1) \osix\ absorption is observed in several large HVCs known to 
lie at large distances from the Galactic plane (e.g., the Magellanic Stream).
Other excellent candidates for HVCs located in the distant Galactic halo
or Local Group based on their ionization properties or abundances include the 
C\,{\sc iv} HVCs toward Mrk~509 (Sembach et al. 1999) and Complex~C 
(Wakker et al.  1999).
\smallskip
\\
2) \osix\ has been observed in the compact HVC toward 
Ton\,S210 (see Figure~1).  This detection is significant since many of 
the observed properties of compact HVCs are best described if they are 
extragalactic clouds within the Local Group.  The compact HVCs
have small angular sizes ($\sim1\deg$) and are isolated.  Their
velocity dispersion as a group of clouds is minimized in the 
Local Group Standard of Rest, and they have a spatial 
distribution similar to that of dwarf galaxies
in the Local Group (Braun \& Burton 1999).  The compact HVC toward Ton\,S210
also has a
sub-solar metallicity that is consistent with a location outside the 
Galaxy, although a more definitive measure of the metallicity is required
to exclude a Galactic location (Sembach et al. 2001a).
\smallskip
\\
3) High velocity \osix\ absorption is observed along sight lines through
the halo along extragalactic lines of sight (e.g., Sembach et al. 2000).
A much larger sample currently being analyzed in conjunction with the 
Wisconsin group (Savage, Wakker, Richter) indicates that high velocity
\osix\ along extended paths through the halo and Local Group is common.
However, there have not yet been any definitive detections of high velocity
\osix\ observed toward halo stars located within a few kiloparsecs of the 
Galactic plane -- an early study by Zsargo et al. (2001, and work in 
preparation) of 10--12 sight lines shows no evidence of high velocity \osix.
\smallskip

Not all of the high velocity \osix\ is located at large distances from the 
Galactic plane. The kinematics and high ion ratios suggest a Galactic origin
for the high velocity \osix\ wing toward 3C\,273. Furthermore, \osix\ is
observed at intermediate-high velocities in several other large-scale 
structures in both the Milky Way and the Magellanic Clouds: for example, the 
Scutum supershell (Sterling et al. 2001) and supernova remnant 
SNR\,0057--7226 in the direction of HD\,5980 in the SMC (Hoopes et al. 2001).
 
\section{Ionization}

The presence of \osix\ in HVCs indicates that the gas is collisionally 
ionized at temperatures of $\sim10^5-10^6$\,K.  Photoionization by starlight
is unlikely.  Photoionization by the extragalactic ultraviolet background 
may produce some \osix, but not in the quantities observed.  The 
typical \osix\ column densities predicted for large ($D >10-20$ kpc) 
low density ($n \sim 10^{-4}$ cm$^{-2}$) clouds at $z \sim 0$ are of the 
order of 
$10^{13}$ cm$^{-2}$ (see Sembach et al. 1999), an order of magnitude less than 
the typical column densities observed.  The distances of the Magellanic 
Stream and Complex~C are not likely to be large enough to accommodate higher 
columns produced by photoionization.  

The coincidence of neutral (or weakly ionized) gas at velocities similar
to those of many of the \osix\ HVCs suggests that production of the  
\osix\ might involve contact interfaces between 
the cool / warm material and hot ($T~\ge~10^6$\,K) gas.  This hotter
gas may take the form of a pervasive, hot Galactic halo or Local Group medium
with a density $n_H < 10^{-4}$ cm$^{-2}$.  Alternatively, the \osix\ may
trace cooling regions of hot gas structures associated with the assembly
of the Milky Way.
Additional STIS data for the \osix\ HVCs observed by $FUSE$ would help to 
constrain the various ionization mechanisms in the clouds.  

\section{Acknowledgments}
I would like to thank and acknowledge my collaborators on various aspects of 
this work, each of whom have contributed to the results presented in this
paper: Blair Savage, Tim Heckman, Claus Leitherer, Brad Gibson, and Mary 
Putman.  I also thank the $FUSE$ Operations Team
at the Johns Hopkins University for their outstanding efforts in making 
$FUSE$ a superb far-ultraviolet observatory.  Finally, I thank John Mulchaey 
and the other workshop organizers for making this a truly enjoyable and 
memorable meeting.

\end{document}